\documentclass[aps,prd,preprint,showpacs,nofootinbib]{revtex4}
\usepackage{epsfig}
\begin{document}
\title{Associated production of top squarks and charginos at the CERN LHC in NLO SUSY-QCD}

\author{Li Gang Jin, Chong Sheng
Li\footnote {E-mail: csli@pku.edu.cn}, and Jian Jun Liu}

\affiliation{Department of Physics, Peking University, Beijing
100871, P.R. China}
\begin{abstract}
We calculate the next-to-leading order inclusive total cross
sections for the associated production processes $pp\rightarrow
\tilde{t}_i\tilde{\chi}_k^-+X$ in the Minimal Supersymmetric
Standard Model at the CERN LHC. Our results provide the
theoretical predictions for the total cross sections for the above
processes. The NLO QCD corrections in general enhance the leading
order total cross sections significantly, and vastly reduce the
dependence of the total cross sections on the
renormalization/factorization scale, which leads to increased
confidence in predictions based on these results.
\\
\noindent [Keywords: Top squark, Chargino, Supersymmetric QCD]
\end{abstract}


\maketitle

The CERN Large Hadron Collider (LHC), with $\sqrt{S}=14$ TeV and a
luminosity of 100 ${\rm fb^{-1}}$ per year \cite{lhc}, will be a
wonderful machine for discovering new physics. In so many new
physical models, the Minimal Supersymmetric Standard Model (MSSM)
\cite{nilles} is one of the most attractive models for the
theorists and the high energy experimenters, and searching for
supersymmetric (SUSY) particles, as a direct experimental
evidence, is one of the prime objectives at the LHC. Therefore, a
good understanding of theoretical predictions of the cross
sections for the production of the SUSY particles is important.
The cross sections for the production of squarks and gluinos were
calculated at the Born level already many years ago \cite{squark}.
To date, the productions of gluinos and squarks
\cite{beenakker2,beenakker6}, top squarks \cite{beenakker3},
sleptons \cite{beenakker4,baer} and gauginos \cite{beenakker4} at
the hadron colliders in the next-to-leading order (NLO) also have
been studied. And recently, the NLO SUSY-QCD analysis of the
associated production of a gaugino ($\tilde{\chi}$) with a gluino
($\tilde{g}$) at the Tevatron and the LHC has been presented in
Ref.~\cite{berger}.

In this Letter we report the NLO QCD (including SUSY QCD)
calculation of the associated production of top squarks (stops)
and charginos at the LHC. Similar to $pp \rightarrow gb
\rightarrow t H^-$ \cite{charged}, which is expected to be a
dominate process for the charged Higgs boson production at the
LHC, the associated production $pp \rightarrow gb \rightarrow
\tilde{t}_i\tilde{\chi}_k^-$ may be also the dominate process for
single top squark or chargino production at the LHC. This is due
to the following reasons: first, the large top quark mass in stop
mass matrix can lead to strong mixing, and induce large mass
difference between the lighter mass eigenstate and the heavier
one, which means that the phase space for the lighter stop will be
great and benefit its production; second, besides containing a
strong QCD coupling between the incoming partons, this process
also includes an enhanced effect from the Yukawa coupling in the
vertex $b-\tilde{t}_i-\tilde{\chi}_k^-$ of the final states. For
simplicity, in our calculation, we neglect the bottom quark mass
except in the Yukawa coupling. Such approximations are valid in
all diagrams, in which the bottom quark appears as an initial
state parton, according to the simplified
Aivazis-Collins-Olness-Tung (ACOT) scheme \cite{acot}. However, it
was pointed out in Ref.~\cite{nbpdf} that the approximations of
the hard process kinematics and the introduction of conventional
bottom quark densities will give rise to sizable bottom quark mass
and kinematical phase space effects, and may overestimate the
inclusive cross section. Very recently, it is shown in
Ref.~\cite{bpdf1} that the bottom parton approach is still valid
if we choose the factorization scale below the average final state
mass: $\mu_f\sim C m_{\rm av}\equiv C
(m_{\tilde{t}_i}+m_{\tilde{\chi}_k^-})/2$ with $C\sim
(1/4,...,1/3)$. Thus, in this Letter, we choose $\mu_f=m_{\rm
av}/3$ when we use the bottom parton approximations.

The leading order (LO) associated production of stops and
charginos proceeds through the subprocess $g(p_a) b(p_b)
\rightarrow \tilde{t}_i(p_1) \tilde{\chi}_k^-(p_2)$ with an
$s$-channel and a $t$-channel. The LO squared matrix element in
$n=4 -2\epsilon$ dimensions, which has been summed the colors and
spins of the outgoing particles, and averaged over the colors and
spins of the incoming ones, is given by
\begin{eqnarray}
\overline{|M^B_{ik}|}^2 =\frac{g_s^2}{12(1 -\epsilon)}
(|l^{\tilde{t}}_{ik}|^2+|k^{\tilde{t}}_{ik}|^2)
(-\frac{t_2+(1-\epsilon)u_2}{s}
+\frac{ss_{\Delta}-u_1(u_2+2t_2)}{st_1}
+\frac{2t_2m_{\tilde{t}_i}^2}{t_1^2}), \label{tree}
\end{eqnarray}
with
\begin{eqnarray}
&& s=(p_a+p_b)^2, \ \ s_\Delta=s-m_{\tilde{t}_i}^2
-m_{\tilde{\chi}_k^-}^2, \nonumber \\
&& t=(p_a-p_1)^2, \ \ \ t_1=t-m_{\tilde{t}_i}^2, \ \ \
t_2=t-m_{\tilde{\chi}_k^-}^2, \nonumber \\
&& u=(p_a-p_2)^2, \ \  u_1=u- m_{\tilde{t}_i}^2, \ \ u_2=u-
m_{\tilde{\chi}_k^-}^2.
\end{eqnarray}
In Eq.~(\ref{tree}), $l^{\tilde{t}}_{ik}$ and $k^{\tilde{t}}_{ik}$
are the left- and right-handed coupling constants of the vertex
$b-\tilde{t}_i-\tilde{\chi}_k^-$, respectively, and are defined as
follows:
\begin{eqnarray}
l^{\tilde{t}}_{ik} =- gR^{\tilde{t}}_{i1}V_{k1}^{\ast} +
\frac{gm_t}{\sqrt{2}m_W\sin\beta} R_{i2}^{\tilde{t}}V_{k2}^{\ast},
\ \ \ \  k^{\tilde{t}}_{ik} =\frac{gm_b}{\sqrt{2}m_W\cos\beta}
R^{\tilde{t}}_{i1}U_{k2}.
\end{eqnarray}
Here the angle $\beta$ is defined by $\tan \beta\equiv v_2/v_1$,
the ratio of the vacuum expectation values of the two Higgs
doublets. Matrices $U/V$ and $R^{\tilde{t}}$ are the chargino and
top squark transformation matrices from interaction to mass
eigenstates \cite{gunion}, respectively.

The NLO corrections to the cross sections can be separated into
the virtual corrections arising from loop diagrams of colored
particles and the real corrections arising from the radiations of
a real gluon or a massless (anti)quark. The virtual corrections
consist of the interference of the LO amplitude $M^B_{ik}$ with
the one-loop amplitudes $M^V_{ik}$ containing the counterterms and
the self-energy, vertex and box diagrams. We carried out the
calculation in t'Hooft-Feynman gauge and used the dimensional
regularization in $n=4 -2\epsilon$ dimensions to regularize the
ultraviolet (UV), soft infrared and collinear divergences in the
virtual loop corrections. However, this method violates the
supersymmetry. In order to restore the supersymmetry the simplest
procedure is through finite shifts in the quark-squark-chargino
couplings \cite{beenakker5}. As for the Dirac matrix $\gamma_5$,
we deal with it using the ``naive" scheme, in which the
$\gamma_5$-matrix anticommutes with the other
$\gamma_{\mu}$-matrices. This is a legitimate procedure at the
one-loop level for anomaly-free theories \cite{gamma5}. The QCD
coupling constant $g_s$ is renormalized in the modified minimal
subtraction ($\overline{MS}$) scheme except that the divergences
associated with the top-quark and colored SUSY particle loops are
subtracted at zero momentum \cite{subtract,beenakker2}. The other
renormalization constants are fixed by the on-mass-shell
renormalization scheme \cite{onmass}, and the renormalization
constant of the stop mixing angle is fixed as shown in
Ref.~\cite{theta}. After renormalization, $M^V_{ik}$ is UV-finite,
but it still contains the infrared (IR) divergences which can be
expressed as
\begin{eqnarray}
M^V_{ik}|_{IR} =\frac{\alpha_s}{2\pi} \Gamma(1+\epsilon)
(\frac{4\pi\mu_r^2}{s})^\epsilon (\frac{A_2^V}{\epsilon^2}
+\frac{A_1^V}{\epsilon})M_{ik}^B,
\end{eqnarray}
where
\begin{eqnarray}
&& A_2^V=-\frac{13}{3}, \ \ \ \ \ \ \ A_1^V=-\frac{43}{6}
-\frac{4}{3}\ln\frac{s}{m_{\tilde{t}_i}^2}
+3\ln\frac{-t_1}{m_{\tilde{t}_i}^2} -\frac{1}{3}\ln\frac{-
u_1}{m_{\tilde{t}_i}^2}.
\end{eqnarray}
Here the infrared divergences include the soft infrared
divergences and the collinear infrared divergences.

The real corrections arising from the real gluon emission $g b
\rightarrow \tilde{t}_i \tilde{\chi}^-_k g$ will produce infrared
singularities, which can be either soft or collinear. These
singularities can be conveniently isolated by the two cutoff phase
space slicing method \cite{cutoff}, in which two cutoffs
$\delta_s$ and $\delta_c$ are introduced, and the partonic cross
section of the real gluon emission can be written as
\begin{eqnarray}
\hat{\sigma}_{ik}^{R}= \hat{\sigma}_{ik}^{S}
+\hat{\sigma}_{ik}^{HC}+ \hat{\sigma}_{ik}^{\overline{HC}}.
\end{eqnarray}
Here the hard non-collinear part
$\hat{\sigma}_{ik}^{\overline{HC}}$ is finite and can be
numerically computed using standard Monte-Carlo integration
techniques \cite{monte}. The soft part $\hat{\sigma}_{ik}^{S}$
contains all the soft infrared divergences, and is given by
\begin{eqnarray}
&&\hat{\sigma}^S_{ik} =\hat{\sigma}^B [\frac{\alpha_s}{2\pi}
\frac{\Gamma(1-\epsilon)}{\Gamma(1-2\epsilon)}
(\frac{4\pi\mu_r^2}{s})^\epsilon] (\frac{A_2^s}{\epsilon^2}
+\frac{A_1^s}{\epsilon} +A_0^s)
\end{eqnarray}
with
\begin{eqnarray}
&& A_2^s=\frac{13}{3}, \hspace{3.0cm} A_1^s= -2A_2^s\ln\delta_s
+\frac{4}{3} +\frac{1}{3} \ln\frac{-u_1}{m_{\tilde{t}_i}^2} -3
\ln\frac{-t_1}{m_{\tilde{t}_i}^2}
+\frac{4}{3}\ln\frac{s}{m_{\tilde{t}_i}^2}, \nonumber
\\
&& A_0^s=2 A_2^s\ln^2\delta_s -2A_1^s \ln\delta_s +
(\frac{4}{3}\frac{\gamma}{\beta} -\frac{1}{4}) \ln\frac{\gamma
+\beta}{\gamma -\beta} +\frac{1}{2}\ln^2\frac{s(\beta
-\gamma)}{t_1} + {\rm Li}_2[1 + \frac{t_1}{s(\gamma -\beta)}]
\nonumber
\\
&& \hspace{1.0cm} -{\rm Li}_2[1 +\frac{s(\gamma +\beta)}{t_1}]
+\frac{1}{6} \{\ln^2\frac{s(\beta - \gamma)}{u_1}
-\frac{1}{2}\ln^2\frac{\gamma +\beta}{\gamma -\beta} +2{\rm
Li}_2[\frac{t_1 +s (\beta +\gamma)}{s(\beta -\gamma)}] \nonumber
\\
&& \hspace{1.0cm} -2{\rm Li}_2[\frac{t_1 +s
(\beta-\gamma)}{-u_1}]\},
\end{eqnarray}
where $\gamma=(s + m_{\tilde{t}_i}^2
-m_{\tilde{\chi}_k^-}^2)/(2s)$ and
$\beta=\sqrt{\gamma^2-m_{\tilde{t}_i}^2/s}$. The hard collinear
part $\hat{\sigma}_{ik}^{HC}$ contains the collinear divergences,
and is given by \cite{cutoff}
\begin{eqnarray}
&& d\sigma^{HC}_{ik} =d\hat{\sigma}^B_{ik} [\frac{\alpha_s}{2\pi}
\frac{\Gamma(1-\epsilon)} {\Gamma(1-2\epsilon)}
(\frac{4\pi\mu_r^2}{s})^\epsilon] (-\frac{1}{\epsilon})
\delta_c^{-\epsilon} [P_{bb}(z,\epsilon)G_{b/p}(x_1/z)G_{g/p}(x_2)
\nonumber
\\ && \hspace{1.4cm} + P_{gg}
(z,\epsilon)G_{g/p}(x_1/z) G_{b/p}(x_2) +(x_1\leftrightarrow x_2)]
\frac{dz}{z} (\frac{1 -z}{z})^{-\epsilon} dx_1 dx_2,
\end{eqnarray}
where $G_{b(g)/p}(x)$ is the bare parton distribution function
(PDF), and the unregulated splitting functions
$P_{ij}(z,\epsilon)$ can be related to the Altarelli-Parisi
splitting kernels \cite{altarelli1} as
$P_{ij}(z,\epsilon)=P_{ij}(z) +\epsilon P_{ij}'(z)$, explicitly
\begin{eqnarray}
&& P_{qq}(z)=C_F \frac{1 +z^2}{1-z}, \hspace{4.5cm} P_{qq}'(z)=
-C_F (1-z), \nonumber \\
&& P_{gg}(z) =2 N [\frac{z}{ 1-z} +\frac{1-z}{z} +z(1-z)],
\hspace{1.0cm} P_{gg}'(z)=0
\end{eqnarray}
with $C_F=4/3$ and $N=3$.

In addition to the real gluon emission, other real emission
corrections to the inclusive cross section for $pp\rightarrow
\tilde{t}_i \tilde{\chi}^-_k$ at NLO involve the processes with an
additional massless (anti)quark in the final states ($
q=u,d,s,c$):
\begin{eqnarray}
&& g + g \rightarrow \tilde{t}_i+ \tilde{\chi}_k^-
 + \bar{b}, \label{gg} \\
&& q/\bar{q} + b \rightarrow \tilde{t}_i +
\tilde{\chi}_k^-  + q/\bar{q}, \label{qb} \\
&& b/\bar{b} +b \rightarrow \tilde{t}_i  +
\tilde{\chi}_k^-  + b/\bar{b}, \label{bb} \\
&& q + \bar{q} \rightarrow \tilde{t}_i + \tilde{\chi}_k^- +
\bar{b} \label{qbq}.
\end{eqnarray}
Since the contributions from the processes (\ref{gg})-(\ref{bb})
contain the initial state collinear singularities, we also need to
use the two cutoff phase space slicing method \cite{cutoff}.
However, we only split the phase space into two regions, because
there are no soft divergences here. Thus, according to the
approach shown in Ref.~\cite{cutoff}, the cross sections for the
processes with an additional massless quark in the final states
can be expressed as ($ q=u,d,s,c,b$)
\begin{eqnarray}
&& d\sigma_{ik}^{add}= \sum_{(\alpha,\beta)}
\hat{\sigma}_{ik}^{\overline{C}}(\alpha\beta\rightarrow
\tilde{t}_i \tilde{\chi}_k^- +X) [G_{\alpha/p}(x_1)
G_{\beta/p}(x_2) +(x_1\leftrightarrow x_2)] dx_1dx_2 \nonumber
\\&& \hspace{1.0cm}
+d\hat{\sigma}^B_{ik} [\frac{\alpha_s}{2\pi}
\frac{\Gamma(1-\epsilon)} {\Gamma(1-2\epsilon)}
(\frac{4\pi\mu^2_r}{s})^\epsilon] (-\frac{1}{\epsilon})
\delta_c^{-\epsilon} [P_{bg}(z,\epsilon)G_{g/p}(x_1/z)G_{g/p}(x_2)
\nonumber
\\ && \hspace{1.0cm} + \sum_{\alpha=q,\bar{q}} P_{g\alpha}
(z,\epsilon)G_{\alpha/p}(x_1/z) G_{b/p}(x_2) +(x_1\leftrightarrow
x_2)] \frac{dz}{z} (\frac{1 -z}{z})^{-\epsilon} dx_1 dx_2,
\label{add}
\end{eqnarray}
where
\begin{eqnarray}
&& P_{gq}(z) =C_F \frac{1 +(1-z)^2}{z}, \hspace{2.0cm} P_{gq}'(z)=
-C_F z, \nonumber
\\ &&
P_{qg}(z) =\frac{1}{2}[z^2 +(1-z)^2], \hspace{2.0cm}
P_{qg}'(z)=-z(1-z).
\end{eqnarray}
The first term in Eq.(\ref{add}) represents the non-collinear
cross sections for the four processes. Moreover, for the
subprocesses $gg/q\bar{q} \rightarrow
\tilde{t}_i\bar{\tilde{t}}_i^\ast \rightarrow
\tilde{t}_i\tilde{\chi}_k^-\bar{b}$ ($q=u,d,c,s,b$), assuming
$m_{\tilde{t}_i}
> m_{\tilde{\chi}_k^-}$, the stop momentum can approach the
$m_{\tilde{t}_i}$ mass shell, which will lead to singularity
arising from the stop propagator. Following the analysis shown in
Ref.~\cite{beenakker2}, this problem can easily be solved by
introducing the non-zero stop widths $\Gamma_{\tilde{t}_i}$ and
regularizing in this way the higher-order amplitudes. However,
these on-shell stop contributions are already accounted for by the
LO stop pair production with a subsequent decay into a chargino
and a bottom quark, and thus should not be considered as a genuine
higher order correction to the associated production of top
squarks and charginos. Therefore, to avoid double counting, these
pole contributions will be subtracted in our numerical
calculations below in the same way as shown in Appendix B of
Ref.~\cite{beenakker2}.

After adding the renormalized virtual and real corrections, the
partonic cross sections still contain the collinear divergences,
which can be absorbed into the redefinition of the PDF at NLO, in
general called mass factorization \cite{altarelli}. This procedure
in practice means that first we convolute the partonic cross
section with the bare PDF $G_{\alpha/p}(x)$, and then use the
renormalized PDF $G_{\alpha/p}(x,\mu_f)$ to replace
$G_{\alpha/p}(x)$. In the $\overline{\rm MS}$ convention, the
scale dependent PDF $G_{\alpha/p}(x,\mu_f)$ is given by
\cite{cutoff}
\begin{eqnarray}
G_{\alpha/p}(x,\mu_f)= G_{\alpha/p}(x)+
\sum_{\beta}(-\frac{1}{\epsilon}) [\frac{\alpha_s}{2\pi}
\frac{\Gamma(1 -\epsilon)}{\Gamma(1 -2\epsilon)} (\frac{4\pi
\mu_r^2}{\mu_f^2})^\epsilon] \int_x^1 \frac{dz}{z} P_{\alpha\beta}
(z) G_{\beta/p}(x/z).
\end{eqnarray}
This replacement will produce a collinear singular counterterm,
which is combined with the hard collinear contributions to result
in, as the definition in Ref.~\cite{cutoff}, the ${\cal O}
(\alpha_s)$ expression for the remaining collinear contribution:
\begin{eqnarray}
&& d\sigma_{ik}^{coll}=d\hat{\sigma}^B_{ik}[\frac{\alpha_s}{2\pi}
\frac{\Gamma(1-\epsilon)} {\Gamma(1-2\epsilon)}
(\frac{4\pi\mu^2_r}{s})^\epsilon] \{\tilde{G}_{g/p}(x_1,\mu_f)
G_{b/p}(x_2,\mu_f) + G_{g/p}(x_1,\mu_f) \tilde{G}_{b/p}(x_2,\mu_f)
\nonumber
\\ && \hspace{1.2cm}
+\sum_{\alpha=b,g}[\frac{A_1^{sc}(\alpha\rightarrow \alpha
g)}{\epsilon} +A_0^{sc}(\alpha\rightarrow \alpha
g)]G_{g/p}(x_1,\mu_f) G_{b/p}(x_2,\mu_f)
\nonumber
\\ && \hspace{1.2cm}
+(x_1\leftrightarrow x_2)\} dx_1dx_2
\end{eqnarray}
where
\begin{eqnarray}
&& A_1^{sc}(q\rightarrow qg)=C_F(2\ln\delta_s +3/2), \\
&& A_1^{sc}(g\rightarrow gg)=2N\ln\delta_s +(11N -2n_f)/6, \\
&& A_0^{sc}=A_1^{sc}\ln(\frac{s}{\mu_f^2}), \\
&& \tilde{G}_{\alpha/p}(x,\mu_f)=\sum_{\beta}\int_x^{1-
\delta_s\delta_{\alpha\beta}} \frac{dy}{y}
G_{\beta/p}(x/y,\mu_f)\tilde{P}_{\alpha\beta}(y)
\end{eqnarray}
with
\begin{eqnarray}
\tilde{P}_{\alpha\beta}(y)=P_{\alpha\beta} \ln(\delta_c
\frac{1-y}{y} \frac{s}{\mu_f^2}) -P_{\alpha\beta}'(y).
\end{eqnarray}
Finally, the NLO total cross section for $pp\rightarrow
\tilde{t}_i\tilde{\chi}_k^-$ in the $\overline{MS}$ factorization
scheme is
\begin{eqnarray}
&& \sigma^{NLO}_{ik}= \int \{dx_1dx_2
[G_{b/p}(x_1,\mu_f)G_{g/p}(x_2,\mu_f)+ x_1\leftrightarrow
x_2](\hat{\sigma}^{B}_{ik} + \hat{\sigma}^{V}_{ik}+
\hat{\sigma}^{S}_{ik} +\hat{\sigma}^{\overline{HC}}_{ik})
+d\sigma_{ik}^{coll}\} \nonumber
\\ && \hspace{1.0cm} +\sum_{(\alpha,\beta)}\int dx_1dx_2
[G_{\alpha/p}(x_1,\mu_f) G_{\beta/p}(x_2,\mu_f)
+(x_1\leftrightarrow x_2)]
\hat{\sigma}_{ik}^{\overline{C}}(\alpha\beta\rightarrow
\tilde{t}_i \tilde{\chi}_k^- +X).
\end{eqnarray}
Note that the above expression contains no singularities since
$A_2^V +A_2^s =0$ and $A_1^V +A_1^s +A_1^{sc}(b\rightarrow bg)
+A_1^{sc}(g\rightarrow gg) =0$. The explicit expressions of
$\hat{\sigma}^{V}_{ik}$, $\hat{\sigma}^{\overline{HC}}_{ik}$ and
$\hat{\sigma}_{ik}^{\overline{C}}$ have been given in
Ref.~\cite{jin}.

We now present some typical numerical results for total cross
sections for the associated production of top squarks and
charginos at the LHC. In our numerical calculations the Standard
Model (SM) parameters were taken to be $\alpha_{ew}(m_W)=1/128$,
$m_W=80.419$ GeV, $m_Z=91.1882$ GeV, and $m_t=174.3$ GeV
\cite{SM}. We use the two-loop evolution of $\alpha_s(\mu)$
\cite{runningalphas} ($\alpha_s(M_Z)=0.118$), and CTEQ6M (CTEQ6L)
PDFs \cite{CTEQ} throughout the calculations of the NLO (LO) cross
sections. Moreover, in order to improve the perturbative
calculations, we take the running mass $m_b(Q)$ evaluated by the
NLO formula \cite{runningmb} with $m_b(m_b)=4.25$ GeV \cite{mb},
and make the following replacement in the tree-level couplings
\cite{runningmb}:
\begin{eqnarray}
m_b(Q) \ \ \rightarrow \ \ \frac{m_b(Q)}{1+\Delta m_b},
\end{eqnarray}
with
\begin{eqnarray}
&& \Delta m_b=\frac{2\alpha_s}{3\pi}m_{\tilde{g}}\mu\tan\beta
I(m_{\tilde{b}_1},m_{\tilde{b}_2},m_{\tilde{g}}) +\frac{g^2
m_t^2}{16\pi^2 m_W^2 \sin2\beta}\mu A_t
I(m_{\tilde{t}_1},m_{\tilde{t}_2},\mu) \nonumber \\
&& \hspace{1.0cm} -\frac{g^2}{16\pi^2}\mu M_2\tan\beta
\sum_{i=1}^2 [(R^{\tilde{t}}_{i1})^2 I(m_{\tilde{t}_i},M_2,\mu) +
\frac{1}{2}(R^{\tilde{b}}_{i1})^2 I(m_{\tilde{b}_i},M_2,\mu)]
\end{eqnarray}
where $A_t$ is the soft SUSY-breaking parameter, $\mu$ is the
Higgs mixing parameter in the superpotential, and
\begin{eqnarray}
I(a,b,c)=\frac{1}{(a^2-b^2)(b^2-c^2)(a^2-c^2)}
(a^2b^2\log\frac{a^2}{b^2} +b^2c^2\log\frac{b^2}{c^2}
+c^2a^2\log\frac{c^2}{a^2}).
\end{eqnarray}
In order to avoid double counting, it is necessary to subtract
these (SUSY-)QCD corrections from the renormalization constant
$\delta m_b$ in the following numerical calculations. In the
calculations of the stop decay width $\Gamma_{\tilde{t}_i}$, the
two-loop leading-log relations \cite{twoloop} of the neutral Higgs
boson masses and mixing angles in the MSSM were used, and the
CP-odd Higgs boson mass $m_{A^0}$ was fixed to 200 GeV. For the
charged Higgs boson mass the tree-level formula was used.  Other
MSSM parameters are determined as follows: (i) For the parameters
$M_1$, $M_2$ and $\mu$ in the chargino and neutralino matrices, we
take $M_2$ and $\mu$ as the input parameters, and assuming gaugino
mass unification we take $M_1=(5/3)\tan^2\theta_W M_2$ and
$m_{\tilde{g}}= (\alpha_s(m_{\tilde{g}})/\alpha_2)M_2$
\cite{Hidaka}. (ii) For the parameters in squark mass matrices, we
assume $M_{\tilde Q}=M_{\tilde U}=M_{\tilde D}$ and $A_t=A_b=300$
GeV to simplify the calculations. Actually, the numerical results
are not very sensitive to $A_{t(b)}$. Moreover, except in Fig.2,
we always choose the renormalization scale $\mu_r=m_{\rm av}$, and
the factorization scale $\mu_f$ is fixed to $m_{\rm av}/3$.

Fig.1 shows the $m_{\tilde{t}_1}$ dependence of the LO and NLO
predictions for $pp\rightarrow \tilde{t}_i\tilde{\chi}_k^-$,
assuming $\mu=-200$ GeV, $M_2=300$ GeV and $\tan\beta=30$, which
means that two chargino masses are about $182$ GeV and $331$ GeV,
respectively, and $m_{\tilde{t}_2}$ increases from 342 GeV to 683
GeV for $m_{\tilde{t}_1}$ in the range 100--600 GeV. One finds
that the total cross section for the $\tilde{t}_2\tilde{\chi}_2^-$
channel is always smallest and less than 3 fb, but the total cross
sections for other channels are large and range between $10$ fb
and several hundred fb for most values of $m_{\tilde{t}_1}$.
Especially for the $\tilde{t}_1\tilde{\chi}_1^-$ channel, the
total cross section can reach 1 pb for small values of
$m_{\tilde{t}_1}$ (100 GeV $< m_{\tilde{t}_1}<$ 160 GeV), which is
almost the same as ones of top quark and charged Higgs boson
associated production at the LHC. However, when $m_{\tilde{t}_1}$
gets larger and close to $m_{\tilde{t}_2}$, the total cross
section for the $\tilde{t}_2\tilde{\chi}_1^-$ channel is the
largest. Moreover, Fig.1 also shows that the NLO QCD corrections
enhance the LO results significantly, which are in general a few
ten percent. The associated production of $\tilde{t}_1$ and
$\tilde{\chi}_1^-$ is the most important since the total cross
sections are the largest for $m_{\tilde{t}_1}<400$ GeV.  Thus we
mainly discuss this channel below.

In Fig.2 we show the dependence of the total cross sections for
the $\tilde{t}_1\tilde{\chi}_1^-$ production on the
renormalization/factorization scale, assuming $\mu=-200$ GeV,
$M_2=300$ GeV, $\tan\beta=30$, $m_{\tilde{t}_1}=250$ GeV, and
$\mu_r=\mu_f$. One finds that the NLO QCD corrections reduce the
dependence significantly. The cross sections vary by $\pm 15\%$ at
LO but only by $\pm 4\%$ at NLO in the region $0.5<\mu_f/m_{\rm
av}<2.0$. Thus the reliability of the NLO predictions has a
substantial improvement.

The cross sections of the $\tilde{t}_1\tilde{\chi}_1^-$ production
as a function of $\tan\beta$ are displayed for
$m_{\tilde{t}_1}=200, 300$ and 400 GeV in Fig.3, assuming
$\mu=-200$ GeV and $M_2=300$ GeV. From Fig.3, one can see that the
cross sections become large when $\tan\beta$ gets high or low,
which is due to the fact that for the coupling
$b-\tilde{t}_1-\tilde{\chi}_1^-$ at low $\tan\beta$ the top quark
contribution is enhanced while at high $\tan\beta$ the bottom
quark contribution becomes large. Fig.3 also shows that the NLO
QCD corrections enhance the LO total cross sections, and for
$m_{\tilde{t}_1}=200$ and $300$ GeV, the enhancement is more
significant for the medium values of $\tan\beta$ than for the high
and low ones.

In conclusion, we have calculated the NLO inclusive total cross
sections for the associated production processes $pp\rightarrow
\tilde{t}_i\tilde{\chi}_k^-$ in the MSSM at the LHC. Our
calculations show that the total cross sections for the
$\tilde{t}_1\tilde{\chi}_1^-$ production for the lighter top
squark masses in the region 100 GeV $< m_{\tilde{t}_1}<$ 160 GeV
can reach 1 pb in the favorable parameter space allowed by the
current precise experiments, and besides the above case the total
cross sections generally vary from $10$ fb to several hundred fb
except both $m_{\tilde{t}_1}>$ 500 GeV and the
$\tilde{t}_2\tilde{\chi}_2^-$ production channel, which means that
the LHC may produce abundant events of these processes, and it is
very possible to discover these SUSY particles through the above
processes in the future experiments, if the supersymmetry exists.
Moreover, we find that the NLO QCD corrections in general enhance
the LO total cross sections significantly, and vastly reduce the
dependence of the total cross sections on the
renormalization/factorization scale, which leads to increased
confidence in predictions based on these results.

\begin{acknowledgments}
We would like to thank T. Plehn and C.-P. Yuan for useful
discussions and valuable suggestions. This work was supported in
part by the National Natural Science Foundation of China.
\end{acknowledgments}
%
%


\newpage
\begin{figure}
\vspace{4.0cm} \centerline{\epsfig{file=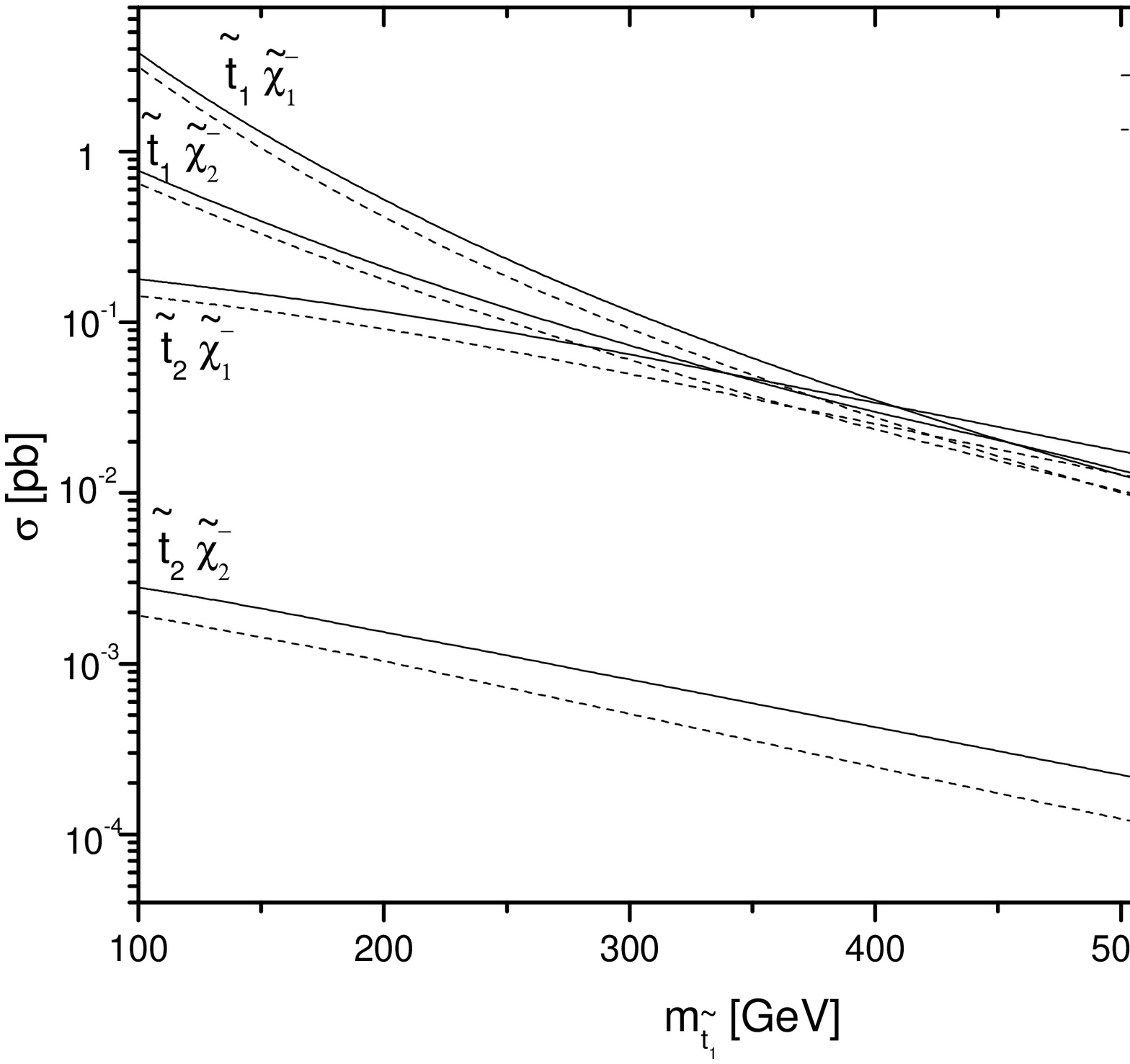, width=500pt}}
\vspace{-4.0cm} \caption[]{Dependence of the total cross sections
on $m_{\tilde{t}_1}$ for the $\tilde{t}_i\tilde{\chi}_k^-$
productions at the LHC, assuming $\mu=-200$ GeV, $M_2=300$ GeV and
$\tan\beta=30$.}
\end{figure}

\begin{figure}
\vspace{4.0cm} \centerline{\epsfig{file=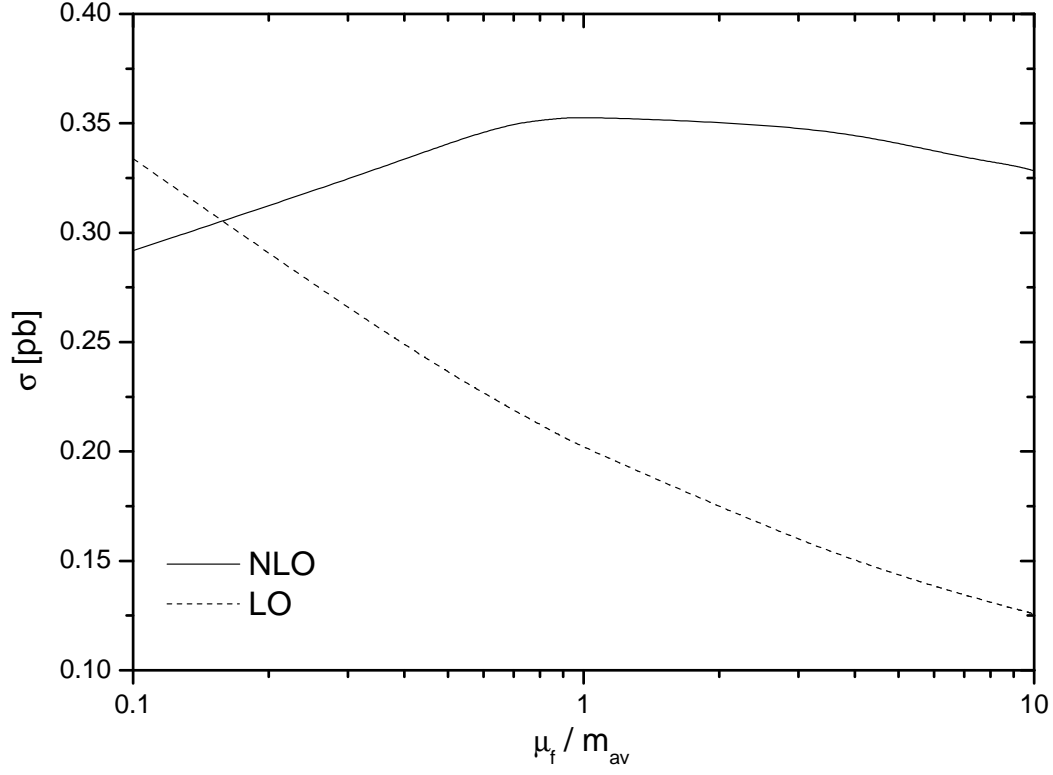, width=500pt}}
\vspace{-4.0cm} \caption[]{Dependence of the total cross sections
for the $\tilde{t}_1\tilde{\chi}_1^-$ production at the LHC on the
renormalization/factorization scale, assuming $\mu=-200$ GeV,
$M_2=300$ GeV, $\tan\beta=30$, $m_{\tilde{t}_1}=250$ GeV,
$\mu_r=\mu_f$ and $m_{\rm av}=(m_{\tilde{t}_1}
+m_{\tilde{\chi}_1^-})/2$.}
\end{figure}

\begin{figure}
\vspace{4.0cm} \centerline{\epsfig{file=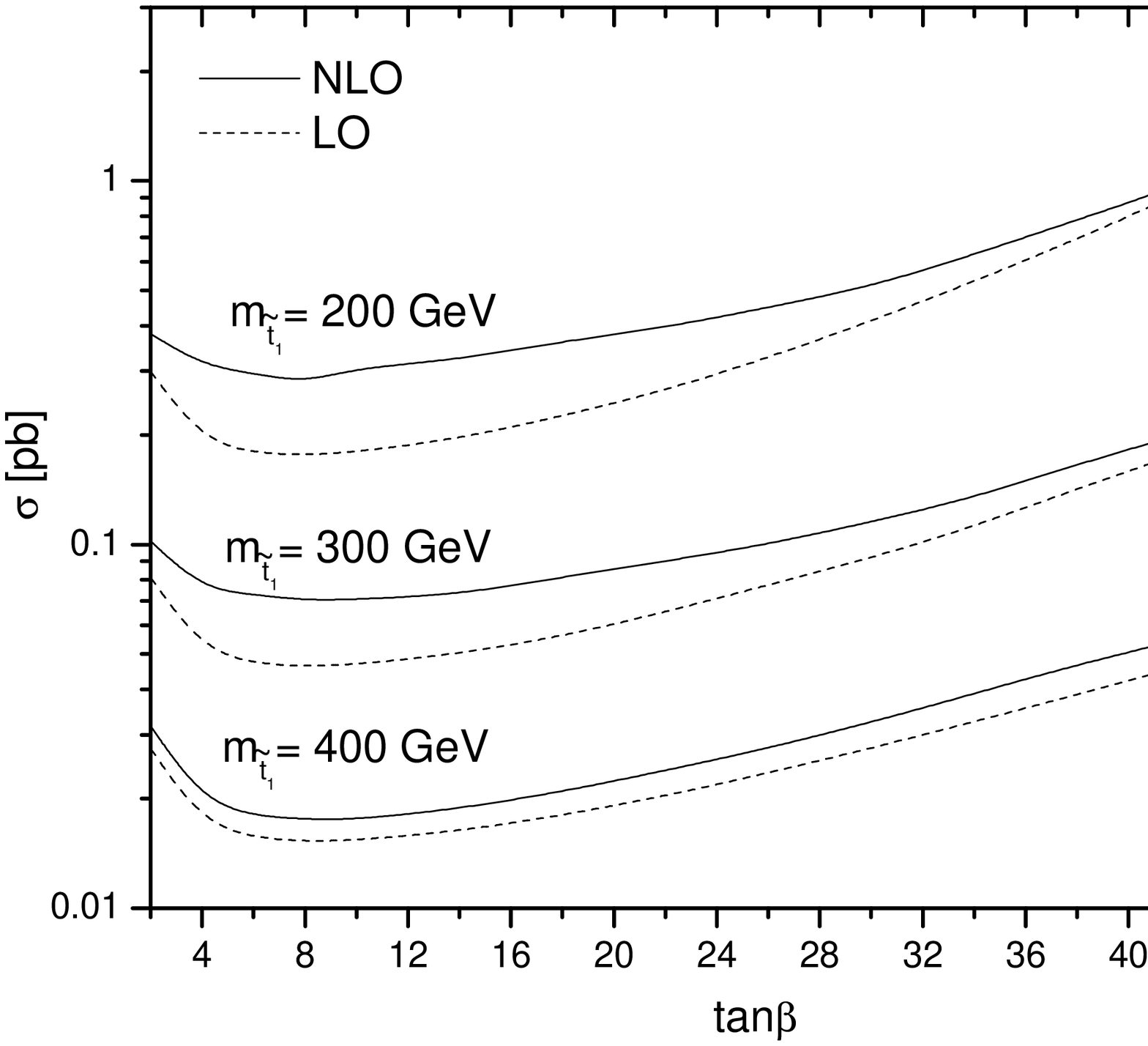, width=500pt}}
\vspace{-4.0cm} \caption[]{Dependence of the total cross sections
for the $\tilde{t}_1\tilde{\chi}_1^-$ production at the LHC on the
parameter $\tan\beta$, assuming $\mu=-200$ GeV and $M_2=300$ GeV.}
\end{figure}


\begin{thebibliography}{}
\bibitem{lhc} F. Gianotti, M.L. Mangano, T. Virdee,
   hep-ph/0204087.
\bibitem{nilles} H.P. Nilles, Phys. Rep. 110 (1984) 1;
   H.E. Haber and G.L. Kane, Phys. Rep. 117 (1985) 75; A.B.
   Lahanas, D.V. Nanopoulos, Phys. Rep. 145 (1987) 1;
   Supersymmetry, Vols. 1 and 2, ed. S. Ferrara (North
   Holland/World Scientific, Singapore, 1987). For
   a recent review, consult S. Dawson, TASI-97 lectures,
   hep-ph/9712464.
\bibitem{squark} G.L. Kane and J.P. Leveille, Phys. Lett. B 112 (1982)
   227; P.R. Harrison and C.H. Llewellyn Smith, Nucl. Phys.
   B 213 (1983) 223; Nucl. Phys. B 223 (1983) 542 (E);
   E. Reya and D.P. Roy, Phys. Rev. D 32 (1985) 645; S.
   Dawson, E. Eichten, C. Quigg, Phys. Rev. D 31 (1985) 1581;
   H. Baer and X. Tata, Phys. Lett. B 160 (1985) 159.
\bibitem{beenakker2} W. Beenakker, R. H$\ddot{\rm o}$pker, M.
   Spira, P.M. Zerwas, Nucl. Phys. B 492 (1997) 51.
\bibitem{beenakker6} W. Beenakker, R. H$\ddot{\rm o}$pker, M.
   Spira, P.M. Zerwas, Phys. Rev. Lett. 74 (1995) 2905;
   Z. Phys. C 69 (1995) 163.
\bibitem{beenakker3} W. Beenakker, M. Kr$\ddot{\rm a}$mer, T. Plehn,
   P.M. Zerwas, Nucl. Phys. B 515 (1998) 3.
\bibitem{beenakker4} W. Beenakker, M. Klasen, M. Kr$\ddot{\rm a}$mer, T. Plehn,
   M. Spira, P.M. Zerwas, Phys. Rev. Lett. 83 (1999) 3780.
\bibitem{baer} H. Baer, B.W. Harris, M.H. Reno, Phys. Rev.
   D 57 (1998) 5871.
\bibitem{berger} E.L. Berger, M. Klasen, T.M.P. Tait, Phys. Lett.
   B 459 (1999) 165; Phys. Rev. D 62 (2000) 095014.
\bibitem{charged} J.F. Gunion, H.E. Haber, F.E. Paige, W.-K. Tung,
   S. Willenbrock, Nucl. Phys. B 294 (1987) 621; R.M.
   Barnett, H.E. Haber, D.E. Soper, Nucl. Phys. B 306 (1988)
   697; F.I. Olness and W.-K. Tung, Nucl. Phys. B 308 (1988)
   813; V. Barger, R.J.N. Phillips, D.P. Roy, Phys.
   Lett. B 324 (1994) 236; C.S. Huang and S.H. Zhu, Phys. Rev. D 60 (1999)
   075012; L.G. Jin, C.S. Li, R.J. Oakes, S.H. Zhu,
   Phys. Rev. D 62 (2000) 053008; S.H. Zhu, hep-ph/0112109.
\bibitem{acot} M. A. Aivazis, J. C. Collins, F. I. Olness, and W.
   K. Tung, Phys. Rev. D 50 (1994) 3102; J. C. Collins,
   Phys. Rev. D 58 (1998) 094002; M. Kramer, F. I. Olness,
   D. E. Soper, Phys. Rev. D 62 (2000) 096007.
\bibitem{nbpdf} D. Cavalli, {\it et al.}, 'The Higgs working group: Summary Report',
   Los Houches 2001, ``Physics at TeV Colliders", hep-ph/0203056; D. Rainwater,
   M. Spira, D. Zeppenfeld, hep-ph/0203187; M. Spira, hep-ph/0211145.
\bibitem{bpdf1} T. Plehn, hep-ph/0206121; F. Maltoni, Z. Sullivan, S. Willenbrock,
   hep-ph/0301033.
\bibitem{gunion} J.F. Gunion and H. Haber, Nucl. Phys. B 272 (1986)
   1.
\bibitem{beenakker5} W. Beenakker, R. H$\ddot{\rm o}$pker, P.M. Zerwas,
   Phys. Lett. B 378 (1996) 159; W. Beenakker, R. H$\ddot{\rm o}$pker,
   T. Plehn, P.M. Zerwas, Z. Phys. C 75 (1997) 349.
\bibitem{gamma5} M. Chanowitz, M. Furman, I. Hinchliffe, Nucl.
   Phys. B 159 (1979) 225.
\bibitem{subtract} J. Collins, F. Wilczek, A. Zee, Phys. Rev.
   D 18 (1978) 242; W. J. Marciano, Phys. Rev.
   D 29 (1984) 580; Phys. Rev.
   D 31 (1984) 213 (E); P. Nason, S. Dawson,
   R.K. Ellis, Nucl. Phys. B 327 (1989) 49; Nucl. Phys. B 335 (1989)
   260 (E).
\bibitem{onmass} A. Sirlin, Phys. Rev. D 22 (1980) 971;
   W. J. Marciano and A. Sirlin, Phys. Rev. D 22 (1980) 2695; Phys. Rev. D
   31 (1985) 213 (E);
   A. Sirlin and W.J. Marciano, Nucl. Phys. B 189 (1981) 442;
   K.I. Aoki et.al., Prog. Theor. Phys. Suppl. 73 (1982) 1.
\bibitem{theta} J. Guasch, W. Hollik, J. Sol$\grave{\rm
   a}$, Phys. Lett. B 437 (1998) 88.
\bibitem{cutoff} B.W. Harris and J.F. Owens, Phys. Rev. D 65 (2002) 094032.
\bibitem{monte} G.P. Lepage, J. Comp. Phys. 27 (1978) 192.
\bibitem{altarelli1} G. Altarelli and G. Parisi, Nucl. Phys.
   B 126 (1977) 298.
\bibitem{altarelli} G. Altarelli, R.K. Ellis, G. Martinelli,
   Nucl. Phys. B 157 (1979) 461; J.C. Collins, D.E. Soper,
   G. Sterman, in: Perturbative Quantum Chromodynamics,
   ed. A.H. Mueller (World Scientific, 1989).
\bibitem{jin} Li Gang Jin, Chong Sheng Li, Jian Jun Liu,
   hep-ph/0210362.
\bibitem{SM} Particle Data Group, D.E. Groom et al, Eur. Phys. J. C 15 (2000)
   1.
\bibitem{runningalphas} S.G. Gorishny, A.L. Kataev, S.A. Larin,
   L.R. Surguladze, Mod. Phys. Lett. A 5 (1990) 2703;
   Phys. Rev. D 43 (1991) 1633; A. Djouadi, M. Spira,
   P.M. Zerwas, Z. Phys. C 70 (1996) 427; A. Djouadi, J.
   Kalinowski, M. Spira, Comput. Phys. Commun. 108 (1998) 56;
   M. Spira, Fortschr. Phys. 46 (1998) 203.
\bibitem{CTEQ} J. Pumplin, D.R. Stump, J. Huston, H.L. Lai, P. Nadolsky, W.K.
   Tung, hep-ph/0201195.
\bibitem{runningmb} M. Carena, D. Garcia, U. Nierste, C.E.M.
   Wagner, Nucl. Phys. B 577 (2000) 88.
\bibitem{mb} M. Beneke and A. Signer, Phys. Lett. B 471 (1999) 233;
   A.H. Hoang, Phys. Rev. D 61 (2000) 034005.
\bibitem{twoloop} M. Carena, M. Quir$\acute{\rm o}$s, C.E.M. Wagner, Nucl. Phys.
   B 461 (1996) 407.
\bibitem{Hidaka} K. Hidaka and A. Bartl, Phys. Lett. B 501 (2001) 78.
\end{thebibliography}
\end{document}